\documentclass[12pt]{iopart}

\usepackage{iopams} 
\usepackage{graphicx} 
\usepackage{color} 
\usepackage{citesort}

\newcommand*{\U}{\mathcal{U}}
\newcommand*{\Heff}{H_\mathrm{eff}}
\newcommand*{\W}{\mathcal{W}}
\newcommand*{\Hrot}{\tilde{H}}
\begin{document}

\title[SSH model with long-range hoppings: topology, driving and disorder]{SSH model with long-range hoppings: topology, driving and disorder}

\author{Beatriz P\'erez-Gonz\'alez$^*$, Miguel Bello$^\dagger$, \'Alvaro G\'omez-Le\'on and 
Gloria Platero}
\ead{$^*$bperez03@ucm.es, $^\dagger$miguel.bello@csic.es}
\address{Instituto de Ciencia de Materiales de Madrid (ICMM-CSIC)}
\vspace{10pt}
\begin{indented}
\item[]January 2018
\end{indented}

\begin{abstract}
The Su-Schrieffer-Heeger (SSH) model describes a finite one-dimensional dimer
lattice with first-neighbour hoppings populated by non-interacting
electrons. In this work we study a generalization of the SSH model including
longer-range hoppings, what we call the extended SSH model. We show that the presence of odd and even hoppings has a very different effect on the topology of the chain. On one hand, even hoppings break particle-hole and sublattice symmetry, making the system topologically trivial, but the Zak phase is still quantized due to the presence of inversion symmetry. On the other hand, odd hoppings allow for phases with a larger topological invariant. This implies that the system supports more edge states in the band's gap. We propose how to engineer those topological phases with a high-frequency driving. Finally, we include a 
numerical analysis on the effect of diagonal and off-diagonal disorder in the 
edge states properties.

\end{abstract}

%\maketitle
%\tableofcontents{}

\section{Introduction}
One of the main tasks condensed matter physics deals with is the understanding of phases of matter.  Traditionally, phase transitions were characterised following Landau's prescription, in terms of an order parameter. Then, the discovery of new phases of matter that did not break any symmetry, nor could be characterised by the usual order parameters, lead to the appearance of topology in condensed matter systems. This new scenario emerged from the merging of physics and topology, and on a more subtle order that lies in the mathematical properties of the electronic wavefunctions. Experimentally, the first developments happened in the study of phase transitions in 2D electronic systems, which displayed a quantised Hall conductance \cite{qhe1982}. Then, the discovery of the fractional QHE \cite{fqhe1983} and of the Spin Hall insulator in HgTe quantum wells \cite{sphi2007} lead to the large variety of systems displaying topological properties that have been discovered so far.

Systems with non-trivial topological properties are changing the way electronics 
is developing. In particular, the discovery of materials with insulating bulk 
and metallic edges, which are also robust under a wide range of perturbations, 
will allow for important advances in spintronics \cite{spintopology2014}, magnetism \cite{magtopology2013} or even further, to 
the development of topological quantum computers \cite{qckitaev2003,majoranabox2017,firstthirdneighbours,generalizedSSH}. Understanding how these 
materials behave in realistic situations is crucial, and the study of the 
classical toy models with new terms is of the utmost relevance. This work is 
framed within this context.  

The starting point of this study is a canonical model of a 1D topological 
insulator: the Su-Schrieffer-Heeger (SSH) model \cite{ssh1980}. It is a tight-binding model for 
non-interacting, spinless electrons confined in a dimer chain. It has been 
extensively studied both theoretically and experimentally \cite{continuumSSH1980,hubbardvspeierls}.

In this work, we analyze the effect of adding arbitrarily long-range hoppings to 
the SSH model, what we call hereafter the extended SSH model. By examining the 
symmetries that are preserved or broken in the resulting system, we can conclude 
that the presence of even and odd hopping terms has different implications on 
the topological properties. Hoppings to even neighbours break particle-hole and 
chiral (also known as sublattice) symmetries, but under certain constraints we 
are able to find gapped configurations with edge states.  On the other hand, odd 
neighbours do not break any fundamental symmetries of the chain, allowing for 
the appearance of larger values of the topological invariant.  More concretely, 
we study in detail the case with first- and second-neighbour hoppings, as well 
as first-, second- and third-neighbour hoppings. 

We also discuss the feasibility of larger winding number configurations
by including an AC driving field. This allows to tune the hopping amplitudes
into unconventional configurations. Furthermore, we examine the effect of diagonal and off-diagonal disorder in the previous results. From the topological point of view, diagonal disorder breaks sublattice symmetry, and therefore affects the topological protection, while off-diagonal disorder maintains this 
symmetry. 

The paper is organized as follows: In section \ref{sec:model} we introduce the 
extended SSH model; in section \ref{sec:topology} we include a characterization 
of its topological properties, considering some relevant concrete examples. 
Section \ref{sec:driving} presents an analysis on the effect of an AC driving 
field on the system, studying several drives with different shapes; in section
\ref{sec:disorder} we study different types of disorder and check their effect
on the edge states of the system. Finally, in section \ref{sec:conclusions} we
present our conclusions.

\section{Extended SSH model}\label{sec:model}

\begin{figure}
	\includegraphics[scale=0.40]{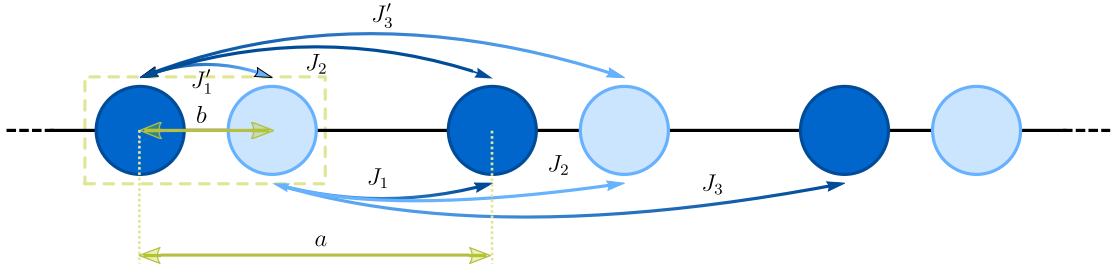}
	\caption{Dimer chain with arbitrarily long-ranged hoppings. For clearness, 
        only hoppings to first, second and third neighbour atoms have been 
        depicted. The unit cell length will be set to $a=1$ hereafter without loss of generality. The intracell parameter is $b$. }
	\label{SSHschematic}
\end{figure}

The Hamiltonian of the dimer chain with hoppings up to $N^{th}$-neighbours 
is given by:
\begin{equation}
    H_N = \sum_{|i-j|\leq N} J_{ij}c^\dagger_i c_j+ \mathrm{H.c.} \,,
    \quad J_{ij} = J^*_{ij} = J(|x_i - x_j|) \,,
    \label{eq:hdimer}
\end{equation}
where $c^\dagger_i$ creates a fermion in the $i^{th}$ site of the chain, and
$J_{ij}=J_{ji}$ is the hopping amplitude connecting the $i^{th}$ and the
$j^{th}$ sites. We can group all the sites in two sublattices $A$ and $B$. All 
the sites with odd indices belong to sublattice $A$ and all the sites with 
even indices belong to sublattice $B$ (see \fref{SSHschematic} for a schematic). 
If we restrict the model to nearest-neighbours only ($N=1$), we recover the 
original SSH model. 

We assume hopping amplitudes are decaying functions of the distance between
sites and define $n = |i-j|$ as the range of the corresponding hopping $J_{ij}$.
Hoppings are denoted as odd or even according to their range. 
It is important to note that in the case of \textit{odd hoppings}, for any 
$n\in \mathbb{N}_\mathrm{odd}$ and site $i$, the $(i+n)^{th}$ and $(i-n)^{th}$
sites are located at different distances. On the contrary, for 
\textit{even hoppings}, all sites are located at the same distance for any $n\in 
\mathbb{N}_\mathrm{even}$. For the sake of simplicity, we will use the following 
notation from now on 
\begin{equation}
	\eqalign{
        J_{2i-n,2i} \equiv J_{n} \,, \quad
        J_{2i,2i+n} \equiv J'_{n} \,,  \quad n\in \mathbb{N}_\mathrm{odd} \\ 
        J_{i,i\pm n} \equiv J_{n} \,, \quad n\in \mathbb{N}_\mathrm{even}}
\end{equation}

For a translationally-invariant system, the Hamiltonian is block-diagonal in the
momentum-space basis. Transforming 
$c_{2j-1} = \frac{1}{\sqrt{M}}\sum_k e^{ikj}a_k$
and $c_{2j} = \frac{1}{\sqrt{M}}\sum_k e^{ikj}b_k$, for $j=1,\dots,M$ ($M$ is 
the number of unit cells in the chain), we can express the Hamiltonian in \Eref{eq:hdimer} with periodic boundary conditions as $H_N =\sum_k\Psi_{k}^{\dagger}\mathcal{H}_N\left(k\right)\Psi_{k}$, 
where we have defined $\Psi_{k}=\left(a_{k},b_{k}\right)^{T}$.
The bulk momentum-space Hamiltonian $\mathcal{H}_{N}\left(k\right)$ is a 
$2\times2$ matrix with the following structure: even hoppings contribute to 
diagonal elements, whereas odd hoppings appear in off-diagonal ones, 
\begin{equation}
	\mathcal{H}_N = \sum_{p} \left(\begin{array}{cc}
		2J_{2p}\cos\left(pk\right) & J'_{2p-1}e^{ik(p-1)}+J_{2p-1}e^{-ikp}\\
		J'_{2p-1}e^{-ik(p-1)}+J_{2p-1}e^{ikp} & 2J_{2p}\cos\left(pk\right)
	\end{array}\right) \,,
	\label{eq:H_k}
\end{equation}
with $p$ ranging from $1$ to $N/2$ if $N$ is even, or $(N+1)/2$ if $N$ is odd. 
$\mathcal{H}_N$ can be written in the basis of the Pauli matrices
$\vec{\sigma}=\{\sigma_{x},\sigma_{y},\sigma_{z}\}$ and the identity
$\mathbf{1}$ as $\mathcal{H}_N=d_{0}(k)\mathbf{1}+\vec{d}(k)\cdot\vec{\sigma}$.
The vector $\vec{d}(k)$ is called the Bloch vector,  and its components are
\begin{eqnarray}
    d_0(k) = \sum_p 2J_{2p}\cos(pk) \,, \\
    d_x(k) = \sum_p \left[J'_{2p-1}\cos\big((p-1)k\big)
                          + J_{2p-1}\cos(pk)\right] \,, \\
    d_y(k) = \sum_p \left[J_{2p-1}\sin(pk)
                          - J'_{2p-1}\sin\big((p-1)k\big)\right] \,, \\
    d_z(k) = 0 \,.
\end{eqnarray}

The dispersion relation takes the form $E_\pm (k) = d_0(k) \pm |\vec{d}(k)|$,
where ``$+$'' and ``$-$'' correspond to the conduction and valence band, 
respectively.  Importantly, the fact that even hoppings of a given range $n$ 
have the same value in both sublattices makes $d_z(k)=0$.

\section{Topology in extended SSH models}\label{sec:topology}
For one-dimensional topological insulators, the topological invariant that 
characterizes different topological phases is the Zak phase $\mathcal{Z}$. 
Equivalently, they can be characterized by the winding of the Bloch vector 
around the origin as $k$ varies across the first Brillouin zone. This quantity 
$\W$ is well-defined only when the Bloch vector lays in a plane containing the 
origin. Both are related to each other as $\mathcal{Z}=\pi \W \ \mathrm{mod} \ 
2\pi$. Owing to the bulk-edge correspondence, the bulk topology manifests itself 
in presence or absence of edge states in a finite system. The number of pairs of 
edge states a system supports corresponds to $|\W|$. The winding number can be 
calculated in terms of the Bloch vector components (see \ref{appendixZ}). 

The Zak phase is a gauge invariant quantity and as such can be measured \cite{measurementZak}. Apart from the SSH model of polyacetylene \cite{ssh1980}, the Zak phase has also been used to characterized linearly conjugated diatomic polymers \cite{conjugatedpolymers}, photonic systems \cite{photonic1,photonic2}, acoustic systems \cite{acousticsystems}, and recently, water wave states \cite{topologicalwater}. 

In the standard SSH model, the winding number can only take two values depending 
on the ratio between first-neighbour hopping amplitudes: $\W = 0$ (trivial 
phase) if $J'_{1}/J_{1}>1$, and $\W = 1$ (non-trivial phase) if $J'_{1}/
J_{1}<1$.  Furthermore, since there are only first-neighbour hoppings in the 
model, it possesses particle-hole symmetry along with time-reversal symmetry and 
chiral (sublattice) symmetry. Therefore, it belongs to the one-dimensional BDI 
class of the Atland-Zirnbauer classification of topological insulators and 
superconductors \cite{tenfoldway}, which admits an infinite number of distinct topological phases. 

\begin{figure}
	\centering
	\includegraphics[scale=0.4]{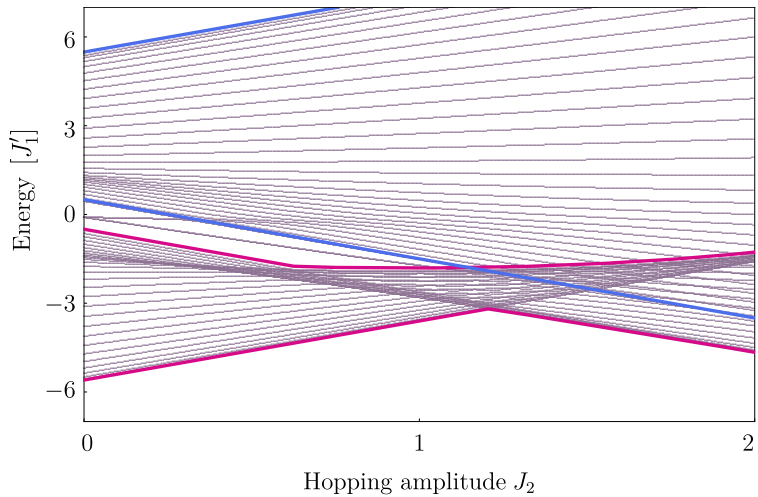}
	\caption{\label{energy13J2} Spectrum for a finite system of $M=30$ and $N=3$, with $J_1=2J'_1$, $J'_3=0.5J'_1$, and $J_3=2J'_1$, as a function of $J_2$. The blue and fuchsia lines represent the maximum and minimum value for the conduction and valence band, respectively. First- and third-neighbour hoppings are chosen such that the system has $\mathcal{W}=2$, i.e., with two pairs of edge states. Second-neighbour alter the energy spectrum, taking the system to a metallic phase because of the overlapping of the two bands. In the gapped phase, note the different behaviour of each pair of edge states. However, the winding number is the same regardless of the value of the hopping amplitude $J_2$. This means that the one-to-one correspondence between $\mathcal{W}$ and the number for edge states is broken.}
\end{figure}

In the extended SSH model the presence of even hoppings breaks particle-hole as 
well as chiral symmetry, changing the system Hamiltonian from BDI class to the 
AI class, which is trivial in 1D. Two clarifications must be made to this 
statement. First, for sufficiently small even hoppings this model supports edge 
states in the band's gap, despite the absence of the aforementioned symmetries.  
Second, even hoppings preserve space-inversion symmetry when chosen as detailed 
in equation \eref{eq:hdimer}, which ensures that the 
winding number is still well-defined. Mathematically, terms proportional to the 
identity matrix (included in $d_{0}$), do not change the eigenstates, and 
therefore the parallel transport, i.e. the Berry connection, is unaffected. 
However, the presence of even hoppings does affect the energy bands and the 
energy levels of a finite system. They may lead to the disappearance of the edge 
states into the bulk bands without the corresponding change in the winding 
number, contrary to the expectation for a true topological phase transition. 
Thus, in general, there is not a one-to-one correspondence between the 
topological invariant and the number of edge states pairs supported by the chain 
as long as even hoppings are present, as shown in \fref{energy13J2}.

Regarding long-range odd hoppings, they preserve all the symmetries of the
standard SSH model, and permit larger values of the topological invariant. 
For a given $N$, the maximum winding number possible is 
$\W_\mathrm{max} = \lfloor (N+1)/2\rfloor$, which is also the maximum number of
pairs of edge states supported by the chain. However, one difficulty for obtaining these phases with larger invariant is that long-range hopping amplitudes must be chosen in a specific way. We will show in next section how we can achieve this by applying ac driving fields. 

In the following lines we will examine in detail two different configurations.

\subsection{First and second neighbour hoppings}

We now study in more detail the effect of even hoppings by considering the case 
of first- and second-neighbour hoppings. As explained before, the study of the 
topology of the system requires the analysis of both bulk and edge properties.  

\subsubsection{Bulk physics \label{bulkphysics}} 
\begin{figure}
	\centering
	\includegraphics[scale=0.4]{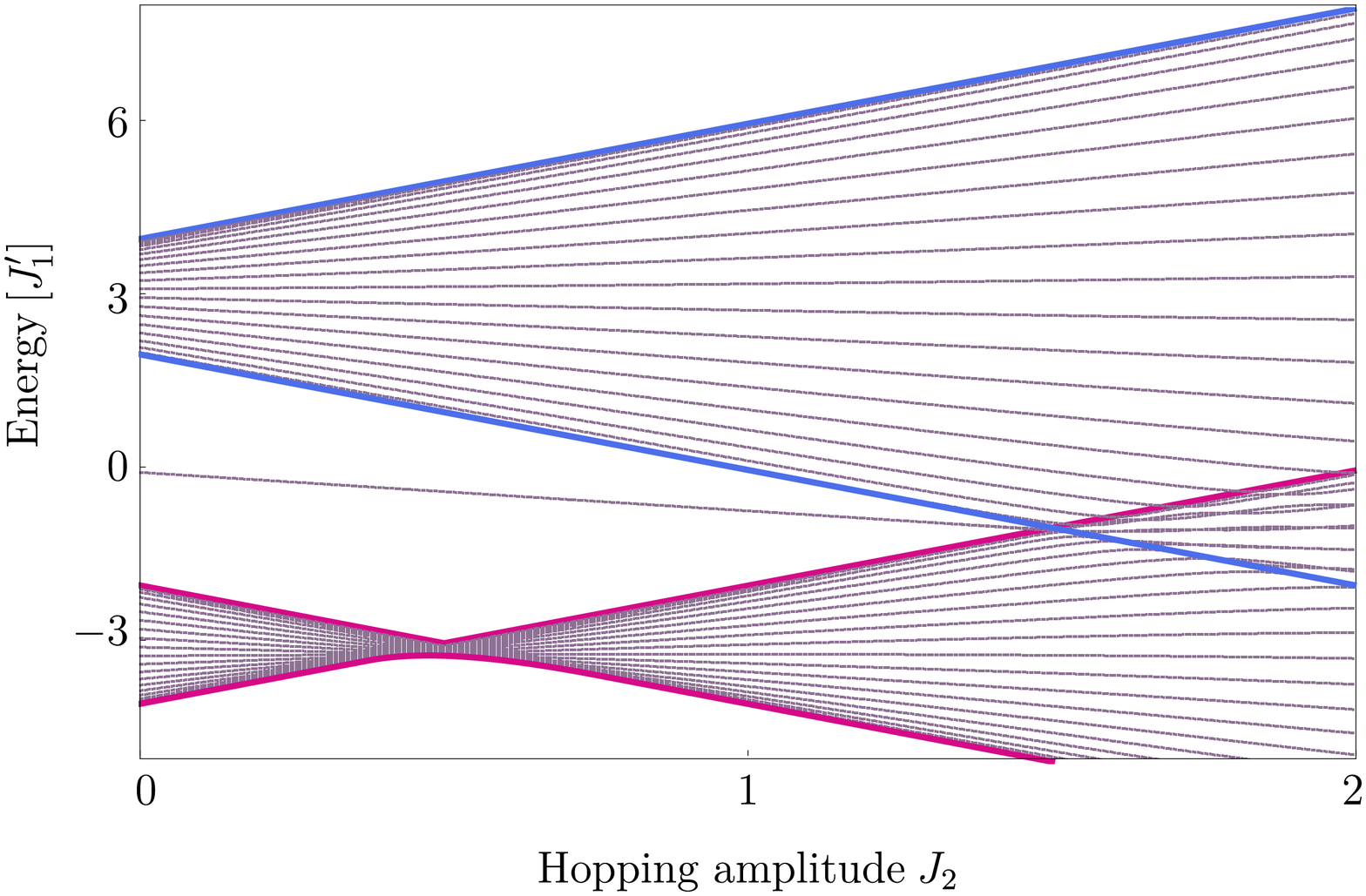}\\
	\vspace{0.7cm}
	\includegraphics[scale=0.4]{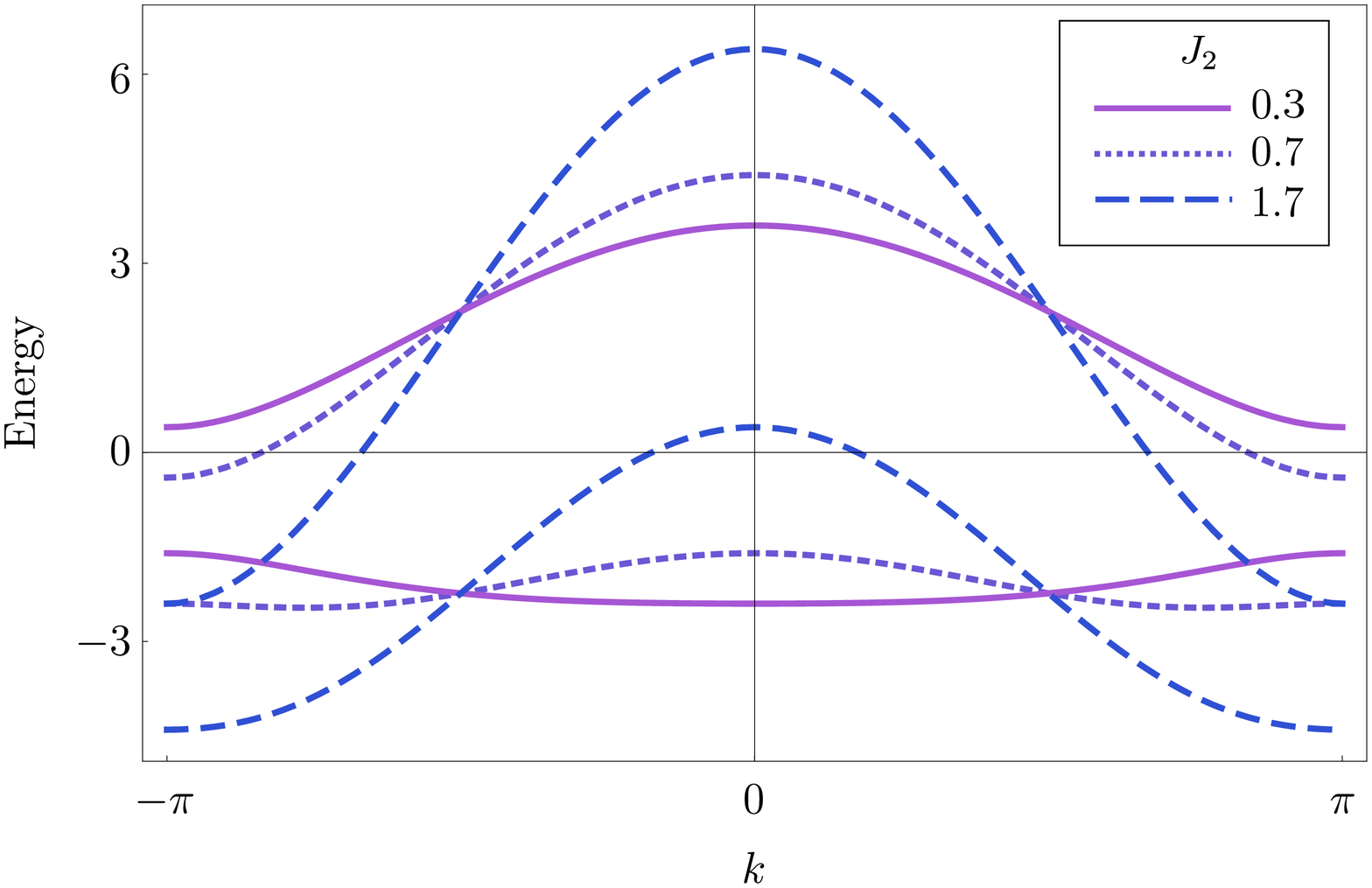}
	\caption{Effect of second-neighbour hoppings on the band structure and energy levels of a finite system.
	
	Top: Spectrum for a finite system of $M=20$ and $N=2$, with $J_{1}=2J'_1$, as a function of $J_{2}$. The blue and fucsia lines represente the maximum and minimum value for the conduction and valence band, respectively. For $J_2<1$, the system has two edge states within the band gap. Their energy decreases as $J_2$ is increased until they penetrate the bulk bands for $J_2\geq J'_1$. Also, the gap goes from direct to indirect at $J_2=0.5J'_1$ (see main text, section \ref{bulkphysics}). In the metallic phase, the bands overlap without crossing.

	Bottom: Bulk band structure for different values of $J_{2}$, given the previous SSH hoppings. Note how particle-hole symmetry is gradually lost as $J_2$ is increased, which is reflected in the loss of symmetry about $E=0$ in the energy spectrum. It is also important to notice that the case with $J_2=0.3J'_1$ has a direct gap, with both the minimum of the conduction band and the maximum of the valence band occur at $k=\pi$. However, for $J_2=0.7J'_1$, the gap is indirect (the minimum of the conduction band occurs at $k=\pm \pi$ and the maximum of the valence band at $k=0$). This corresponds to case (i) discussed in the previous section. On the other hand, 
	 the maximum of the valence band is greater than the minimum of the conduction band for the configuration with $J_2=1.7J'_1$, although the bands never touch. This is an example of band structure of a system in regime (ii), when the system is expected to have metallic properties.}
\label{bulkandedge2}
\end{figure}

The momentum-space Hamiltonian in  \eref{eq:H_k} takes the form 
\begin{equation}
\mathcal{H}_2=\left(\begin{array}{cc}
2J_{2}\cos\left(k\right) & J'_{1}+J_{1}e^{-ik}\\
J'_{1}+J_{1}e^{ik} & 2J_{2}\cos\left(k\right)
\end{array}\right),
\end{equation}
whereas Bloch's vector changes to:
\begin{eqnarray}
d_{0}\left(k\right) & = &2J_{2}\cos\left(k\right)\,, \\
d_{x}\left(k\right) & = & J'_{1}+J_{1}\cos\left(k\right)\,,
\quad d_{y}\left(k\right) = J_{1}\sin\left(k\right)\,, 
\quad d_{z}\left(k\right) = 0 \,
\end{eqnarray}
and the energy dispersion is given by
$E_{\pm}\left(k\right)=2J_{2}\cos\left(k\right)\pm\sqrt{J_{1}^{'2}+J_{1}^{2}+2J'_{1}J_{1}\cos\left(k\right)}$.
This expression makes clear that second-neighbour hoppings break particle-hole
symmetry, which translates into an assymetric band structure about $E=0$. Still,
the specific value of $J_2$ is of utmost importance, as the system properties
change drastically. We can distinguish two regimes (see \fref{bulkandedge2}):
\begin{enumerate}
    \item When $J_{2}<J'_{1}/2$ and $J'_1/J_1>1$ ($\mathcal{W}=0$) or 
        $J_{2}<J_{1}/2$ and $J'_1/J_1<1$ ($\mathcal{W}=1$), the system has insulating properties. This regime corresponds to a gapped phase in which the winding number is still defined by the ratio $J'_1/J_1$ and has a one-to-one correspondence with the number of edge states. It is also significant that the direct gap turns into an indirect gap at $J_{2}=J_{1}/2$ (trivial phase), or $J_{2}=J'_{1}/2$ (topological phase), which means that the minimum energy in the conduction band and the maximum energy in the valence band occur at different values of $k$.  
	\item When $J_{2}\geq J'_{1}/2$ and $J'_1/J_1>1$ or $J_{2}\geq J{}_{1}/2$
    and $J'_1/J_1<1$, the behaviour is expected to be metallic. In this regime the gap is indirect, but the maximum of the valence band (at $k=0$) is equal or greater to the minimum of the conduction band (at $k=\pi$). This means that the energy bands overlap without crossing, which signals the absence of a topological phase transition. 
\end{enumerate}

\subsubsection{Edge physics \label{edgephysics12}}

The topological phase of the SSH chain is characterized by the appearance
of two edge modes. If the thermodynamic limit ($M \rightarrow \infty$), edge states will
be degenerate at $E=0$, each of them being exponentially located
at either the right of left end of the chain. If not, a small splitting
of the order of $(J'_{1}/J_{1})^{N}$ is expected \cite{asboth2016book};
edge states hybridize and become an even and odd superposition of
the states located at one of the ends. The presence of chiral symmetry,
represented by the operator $\mathcal{C}$, ensures that these hybridized
edge states have symmetric energies about $E=0$, since they are chiral
partners of each other: $|\mathrm{edge_{o}}\mathcal{i}=\mathcal{C}|\mathrm{edge}_{\mathrm{e}}\mathcal{i}\rightarrow E_{\mathrm{o}}=-E_{\mathrm{e}}$,  where e and o stands for even and odd parity, respectively.
By solving the dispersion relation we get that edge states have associated
a complex $k=\pi+i\zeta_{\mathrm{SSH}}$, where $\zeta_{\mathrm{SSH}}$
is the inverse of the localization length. The value of $\zeta_{\mathrm{SSH}}$
is a function of the ratio $J'_{1}/J_{1}$ \cite{delplace2011}

\begin{equation}
\frac{J'_{1}}{J_{1}}\approx e^{-\zeta_{\mathrm{SSH}}}\,.
\label{sshdecay}
\end{equation}

\begin{figure}
 	\centering
 	\includegraphics[scale=0.5]{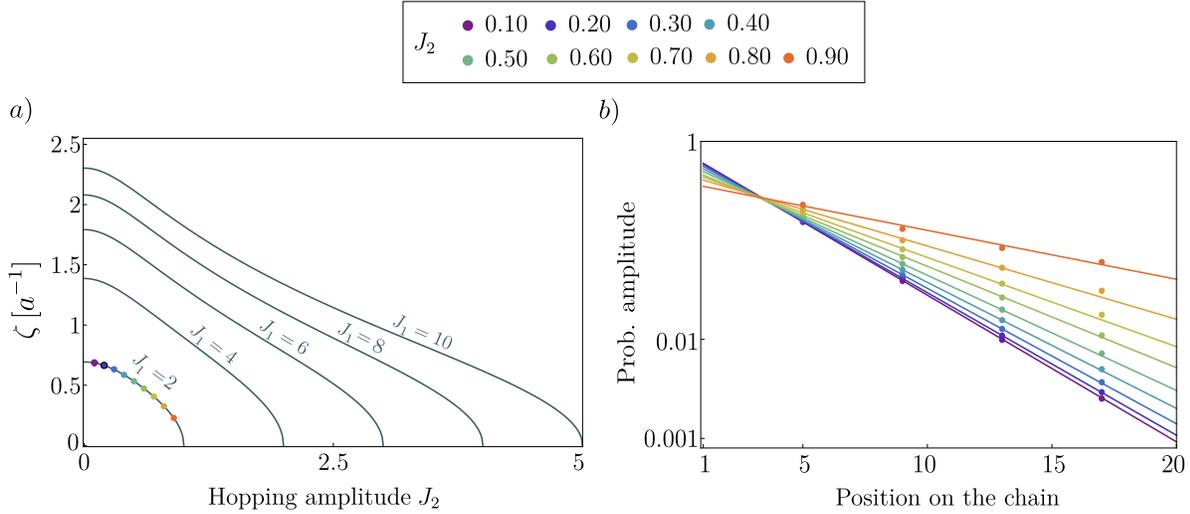}
 	\caption{\label{loclength} a) Localization length of edge states in a chain with first- and second-neighbour hoppings, given fixed $J'_1$, as a function of $J_2$, in units of $J'_1$, see equation \eref{zeta}. For each curve, $\zeta$ goes to zero when $J_2=J_1/2$, which corresponds to the overlapping of the bands. Colored dots in the $J_1=2$ curve correspond to the numerical data obtained by fitting the envelope of the edge states in a finite chain with $M=20$ to a exponential function of the form $\sim e^{-\lambda_{\mathrm{loc}}x}$, for different values of $J_2$ (see legend). \\
 	b) Probability amplitude of edge states wavefunction in logarithmic scale for $J_1=2J'_1$, and $M=20$. Each color corresponds to the values of $J_2$ shown in the legend and in figure \ref{loclength}a. Plotmarkers represent the peak values of the edge states wavefunction, whereas continuous lines depict the numerical fitting to an exponential function. In logarithmic scale, they are represented as lines with slope $-\lambda_{\mathrm{loc}}$. As can be seen, the edge states do decay exponentially into the bulk when second-neighbour hoppings are added.}
 \end{figure}

When second-neighbour hoppings are added, we find the following changes
in the behaviour of the edge states:
\begin{enumerate}
	\item The absence of chiral symmetry implies that $|\mathrm{edge_{o}}\mathcal{i}=\mathcal{C}|\mathrm{edge}_{\mathrm{e}}\mathcal{i}$
	does not hold anymore (see  \fref{sublaticces}).
	\begin{figure}
	\centering
	\includegraphics[scale=0.6]{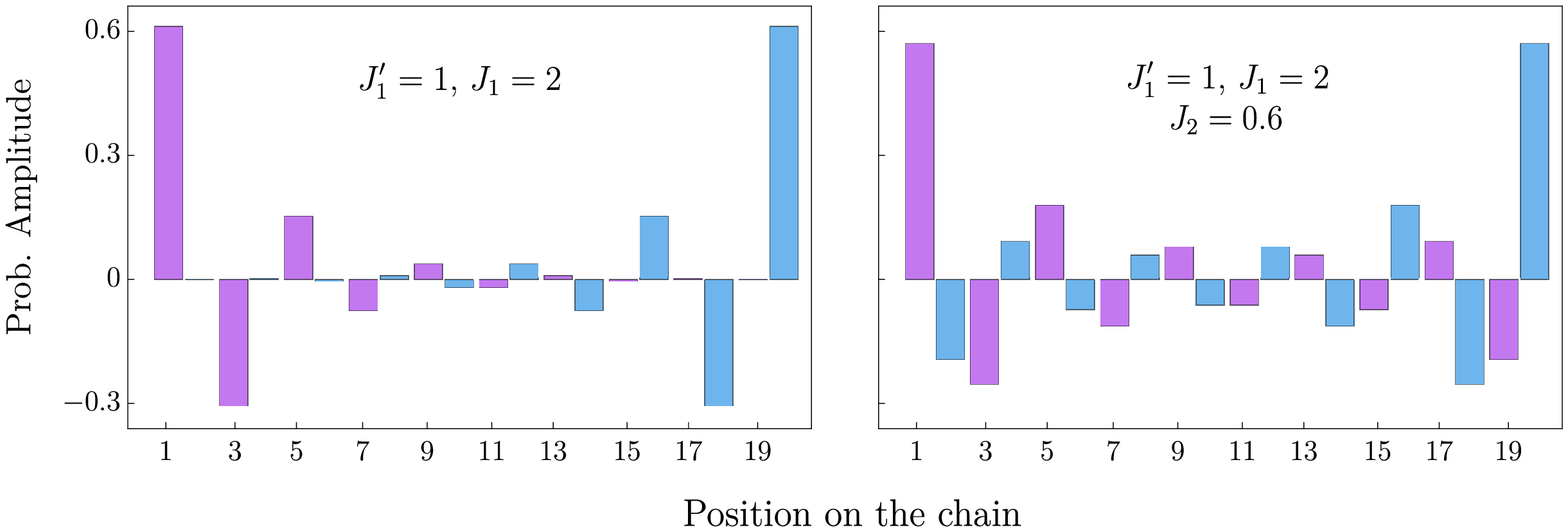}
	\caption{Wave functions of the hybridized edge states (even parity) of a chain with $M=10$ unit cells, with: \\
	Left: only first-neighbour hoppings, $J_{1}=2J'_1$. Edge states in the SSH chain fulfill $|\mathcal{h}\mathrm{edge_{\mathrm{o}}|\mathcal{C}|}\mathrm{edge}_{\mathrm{e}}\mathcal{i}|=1$.  \\
	Right: first- and second-neighbour hoppings, $J_{1}=2J'_1$ and $J_2=0.6J'_1$. The presence of $J_2$ breaks chiral symmetry, and hence $|\mathcal{h}\mathrm{edge_{\mathrm{o}}|\mathcal{C}|}\mathrm{edge}_{\mathrm{e}}\mathcal{i}|=0.8<1$. This quantity becomes smaller as $J_2$ is increased.}
	\label{sublaticces}
	\end{figure}
	
	\item The edge states energy moves away from zero
	as $J_{2}$ increases. Using numerical analysis, we find that the
	energy of both edge states varies linearly with $J_{2}$ according
	to
	\begin{equation}
	E=E_{\mathrm{edge}}-2J_{2}\frac{J'_{1}}{J_{1}}\,,
	\label{energy2neighbours}
	\end{equation}
	where $E_{\mathrm{edge}}$ is the energy of the edge states in the SSH
	chain ($J_2=0$). This expression holds until the enery bands overlap. 
	\item Interestingly, we find that the addition of $J_{2}$ modifies the localization
	length of the edge states, which become less localized as $J_{2}$ is 
    increased.
	First, knowing that the energy of the edge states depends linearly
	on $J_{2}$ as shown in equation \eref{energy2neighbours}, we can
	solve the dispersion relation, obtaining a expression for the $k$
	associated with the edge states in terms of the hopping amplitudes 

\begin{equation}
k=\pm\arccos(\alpha),\,\alpha=-\frac{J'_{1}}{J_{1}}+\frac{J_{1}J'_{1}}{4J_{2}^{2}}-\frac{1}{4J_{2}^{2}}\sqrt{4J_{2}^{2}(J_{1}^{2}-J_{1}^{'2})+J_{1}^{2}J_{1}^{'2}}\,.
\end{equation}

In order for the state to be localized, we search for a solution $k$
of the form $k=\pi\pm i\zeta$, where $\zeta=1/\lambda_{\mathrm{loc}}$.
%Thus, we expect the argument of the arccosine function to be $|\alpha|>1$,
%since any $|\alpha|\leq1$ would result in a real $k$, which corresponds
%to a non-localized state. 
If we rewrite the previous equation as $\cos(k)=\cos(\pi+i\zeta)=-\cosh(\zeta)=\alpha$,
we can give an analytic expression for $\zeta$

\begin{equation}
\zeta=\frac{1}{\lambda_{\mathrm{loc}}}=\mathrm{arccosh}\left(-\alpha\right)\,.
\label{zeta}
\end{equation}

In the limit $J_{2}\rightarrow J_{1}/2$
(when the bands overlap and the edge states penetrate the energy bands),
for which $\zeta\rightarrow0$. In the limit $J_{1}\rightarrow J'_{1}$, i.e. one-dimensional atomic chain,
(when the band gap is closed and the system has metallic behaviour), $\zeta\rightarrow0$
independently of the value of $J_{2}$. In both cases, localized behaviour
is lost, which agrees with the analytic and numerical results previously
obtained. %In the SSH
%chain, $\zeta_{\mathrm{SSH}}$ can be written in the following form

%\begin{equation}
%\zeta_{\mathrm{SSH}}=\mathrm{arccosh}\left(\frac{J^{'2}_{1}+J_{1}^{2}}{2J_{1}J'_{1}}\right)
%\end{equation}

As can be seen in \fref{loclength}, $\zeta$ is affected differently by $J_2$ depending on the value of first-neighbour hopping amplitudes. As $J'_1/J_1$ gets closer to one, that is, as we approach the metallic limit, the presence of $J_2$ has less impact on $\zeta$.
\end{enumerate}

\subsection{First- and third-neighbour hoppings}

When first- and third-neighbour hoppings are considered, the system preserves time-reversal, particle-hole and chiral symmetry, and thus it belongs to the BDI class, just as the standard SSH model. Therefore, the topological invariant is well-defined and there is a one-to-one correspondence between its value and the number of edge states supported by the system.

\begin{figure}
	\centering 
    \includegraphics[width=\linewidth]{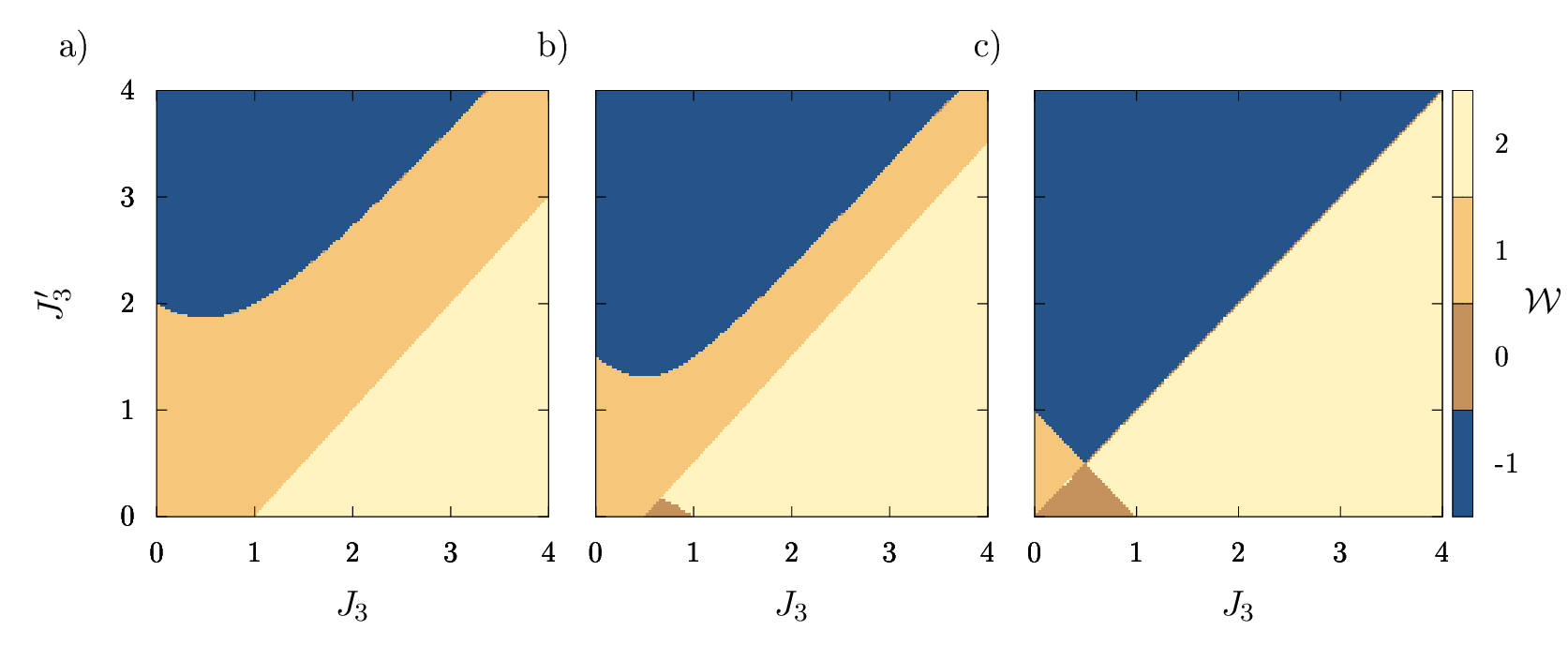}
	\caption{\label{phasemap} Topological phase diagram of as a function of
        third-neighbour hoppings, for different $J'_3$, and $J_3$ (expressed in
        units of $J'_1$). Second neighbours are set to zero in all of them 
            $J_2=0$: 
    a) $J_1=2J'_1$, b) $J_1=1.5J'_1 $, c) $J_1=J'_1$. 
    Figures a) and b) fulfill $J'_1/J_1<1$, which
    corresponds to a SSH topological insulator. Figure c) corresponds to an homogeneous chain ($J'_1=J_1$) which is gapped due to third neighbour hoppings.}
\end{figure}

The Bloch vector has the following non-zero components, $d_x(k)=J'_1+(J_1+J'_3)\cos(k)+J_3\cos(2k)$, and $d_y(k)=(J_1-J'_3)\sin(k)+J_3\sin(2k)$, in terms of which the winding number can be calculated. A topological phase diagram is obtained as a function of $J'_3$ and $J_3$ for different first-neighbour hoppings (see \fref{phasemap}), setting to zero second-neighbour hoppings in order to preserve chiral symmetry. The presence of long-range hoppings enriches the phase map, making possible the existence of configurations with $\mathcal{W}=2$ and $\mathcal{W}=-1$.

\begin{figure}[!htb]
	\centering
	\includegraphics[scale=0.5]{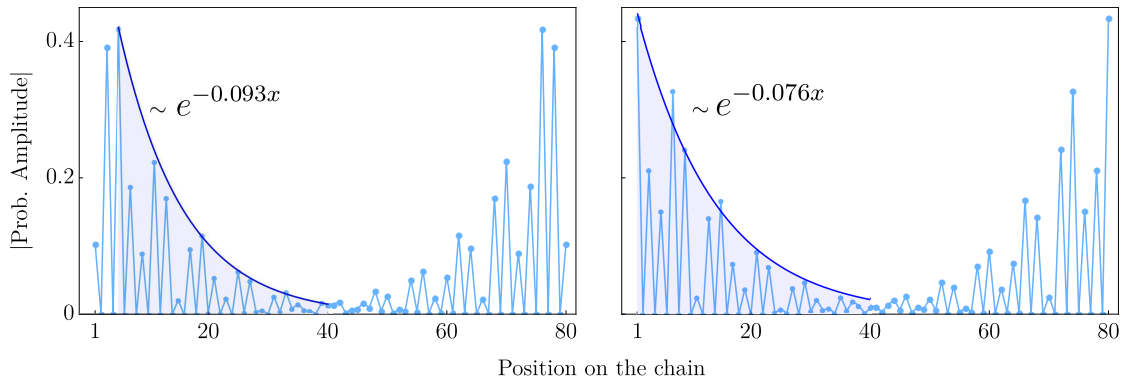}
	\caption{\label{edgestates13} Absolute value of the  edge states wavefunction of a chain with $M=40$ unit cells, and hopping amplitudes $J_1=4J'_1/3$, $J_2=0$, $J'_3=J'_1/5$, and $J_3=J'_1$ ($\mathcal{W}=2$). The continuous, blue line represents the fitting of the envelope to a exponential function. Each edge state depicted belongs to a different pair, and thus the peak of maximum probability occurs in a different site of the chain.}
\end{figure}

Interestingly, dimer chains with $\mathcal{W}=2$ support two pair of edge states. Owing to the presence of chiral symmetry, each pair carries two chiral partners, whose energies are related by $E_{\mathrm{e}}=-E_{\mathrm{o}}$. In the thermodynamic limit, when $M\rightarrow \infty$, these zero modes are located at either the right or left edge of the chain and can be chosen to have support on one of the sublattices, just as those in the SSH model. However, one remarkable distinction from the latter is the fact that each pair has a different spatial dependence, which in turn differ from that of the SSH model edge states. First, the peak of maximum probability amplitude is located at a different site for each pair. Depending on how hopping amplitudes are tuned, pairs can be maximally located at either the first, third, or fifth site of the chain. Moreover, the envelope of the edge states wavefunction decays exponentially into the bulk, but the probability amplitude on each site does not decrease monotonically. The larger the system, the more nicely the envelope fits into a exponential decay (see figure \ref{edgestates13}). 

\section{Periodic driving}\label{sec:driving}

As we have shown, phases with more than a single pair of edge states are
possible, although they require unconventional hopping parameters, such that
hopping amplitudes to further neighbours are larger than those to closer
neighbours. In a regular system, however, one may expect hopping amplitudes
to decrease with increasing distance. One way to overcome this consists in
using a periodic driving, which in the high-frequency regime makes the system
behave as if it were governed by an effective static Hamiltonian, with the 
possibility to change the effective hopping amplitudes by tuning the driving 
parameters \cite{drivenchain2013}. 

With this purpose in mind, we include in the system Hamiltonian $H_N$ a
time-dependent term $H_{AC}(t) = E(t)\sum_{j}x_j n_j$, corresponding to a 
homogeneous ac field $E(t)$ that couples to the charge (or mass) of the 
particles. $E(t)$ is a periodic function of time with period $T=2\pi/\omega$. 
Using a high-frequency expansion, we can derive an effective Hamiltonian 
$H_\mathrm{eff}$ expressed as a power series in $1/\omega$, see appendix 
\ref{app:Floquet}. To lowest order, $H_\mathrm{eff}$ is simply the time 
average of the total Hamiltonian over one period. Thus, the structure of 
the hoppings is maintained, but the hopping amplitudes become renormalized 
as
\begin{equation}
    J_{ij}\rightarrow J_{ij}\frac{1}{T}\int_0^T dt e^{i A(t) d_{ij}} \equiv 
    J_{ij}f(E_0d_{ij}/\omega)\,.
    \label{eq:Jeff}
\end{equation}
Here $A(t)$ is the vector potential corresponding to the ac field
$E(t)=-\partial_t A(t)$ and $d_{ij}=x_i-x_j$ is the distance between the 
$i^{th}$ and $j^{th}$ sites. We will assume that the decay of hopping amplitudes 
with distance is exponential, $J_{ij}=J_0 e^{-d_{ij}/\lambda}$. Below, in table
\ref{tab:renormalizations} we specify three different driving protocols
studied in this work, with the corresponding hopping renormalization they 
produce.
\begin{table}
	\begin{tabular}{c c c}
		\br 
        Driving & Vector potential, $A$
        &  Hopping renormalization, $f$ \tabularnewline
		\hline 
        simple sinusoidal & $-\frac{E_0}{\omega}\sin(\omega t)$
        & $\mathcal{J}_0\left(\frac{E_0 d_{ij}}{\omega}\right)$
        \tabularnewline
        double sinusoidal & $-\frac{E_0}{\omega}
        \left[\sin(\omega t) + \sin(3\omega t)\right]$
        & $\sum_n \mathcal{J}_{-3n}\left(\frac{E_0 d_{ij}}{\omega}\right)
        \mathcal{J}_{n}\left(\frac{E_0 d_{ij}}{\omega}\right)$
        \tabularnewline
        square-wave & $\left\{ \begin{array}{l l l} -E_0 t & \mathrm{if} & 0<t<T/2 \\ 
        E_0(t-T) & \mathrm{if} &  T/2<t<T \end{array}\right.$ &             
        $2i \left(e^{-iE_0Td_{ij}/2} - 1\right)/E_0 T d_{ij}$
        \tabularnewline
		\br
	\end{tabular}
	\caption{Different driving protocols with the 
        corresponding hopping renormalization}
    \label{tab:renormalizations}
\end{table}

\begin{figure}[!htb]
    \centering
    \includegraphics[scale=1.2]{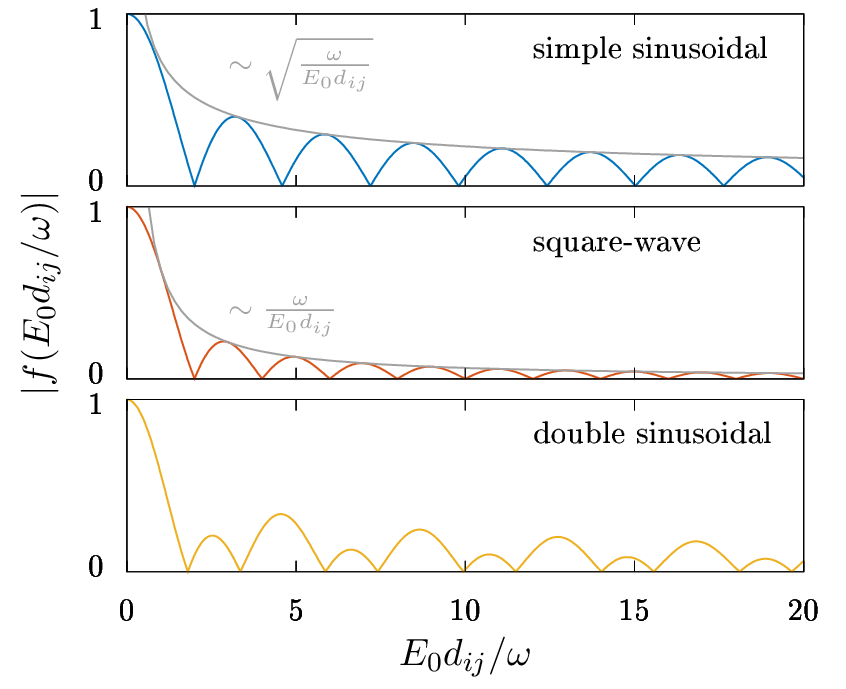}
    \caption{Comparison between the hopping renormalization functions of the
    different drivings studied. Note how the zeros of $f$
for the square-driving are equally spaced, while its envelope (grey
line) decays faster than for the sinusoidal drivings.}
    \label{fig:renormalizations}
\end{figure}

\begin{figure}[!htb]
    \centering
    \includegraphics[width=\linewidth]{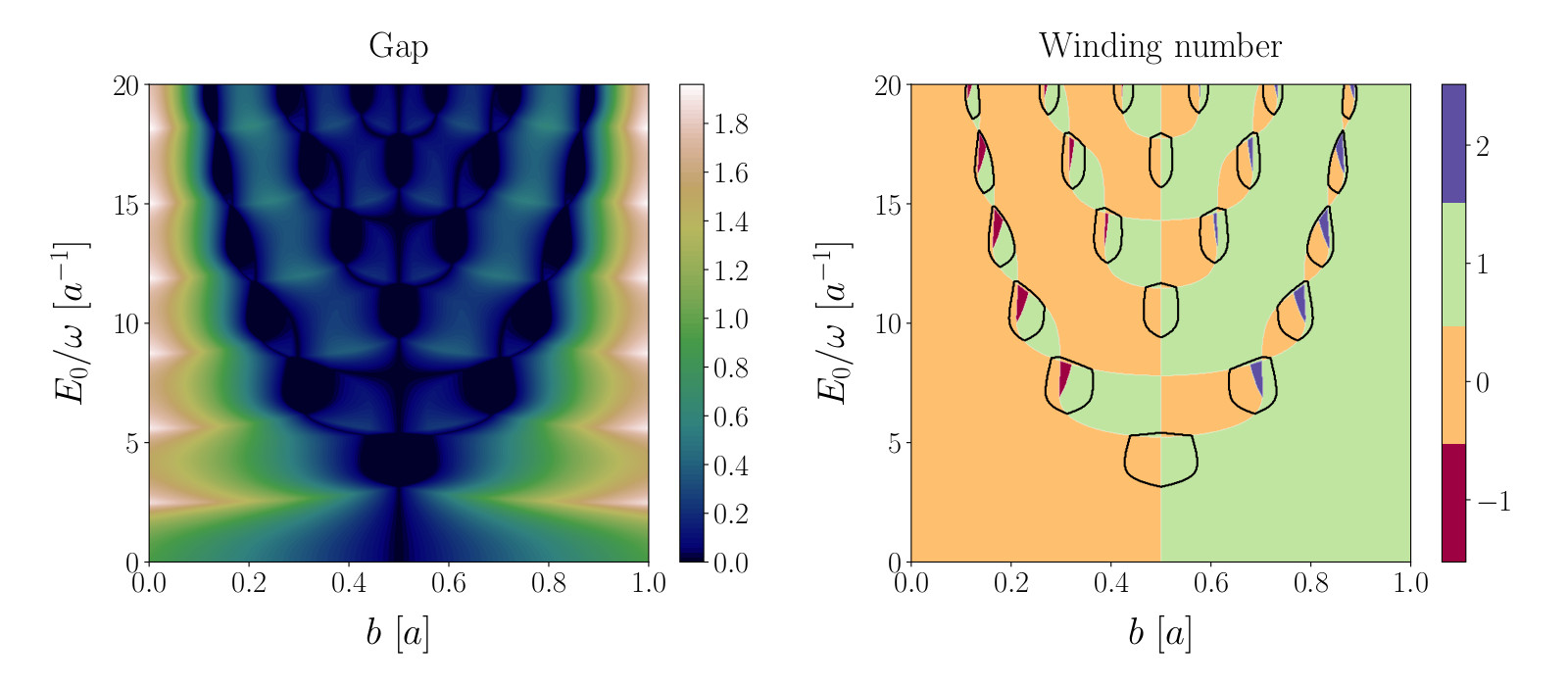}
    \caption{Phase diagram of the extended SSH model with a sinusoidal driving $A(t)=-\frac{E_0}{w}\sin(wt)$. Hopping amplitudes decay exponentially with distance. Chosing $\lambda=a$, only hoppings with range up to $N=10$ have a significant contribution. The gap is expressed in units of $J_0$. In the plot of the winding number, black curves show the contour level where the gap vanishes.}
    \label{fig:PhaseDiagram_driven}
\end{figure}

For a simple sinusoidal drive with amplitude $E_0$ and frequency $\omega$, the 
hopping renormalization is given by the zeroth-order Bessel function of the 
first kind $\mathcal{J}_0(E_0 d_{ij}/\omega)$ \cite{Hanggi1998}.  This allows to 
cancel the hoppings to next-nearest neighbours by tuning $E_0 a/\omega$ to one 
of the zeros of $\mathcal{J}_0$. In this manner, it is possible to recover the 
chiral symmetry in chains with hoppings up to third neighbours. Nonetheless, it 
is impossible to zero out all even hoppings with this driving. Interestingly, we 
obtain winding numbers up to $\W=2$, but only for metallic phases, see figure 
\ref{fig:PhaseDiagram_driven}.  

\begin{figure}[!htb]
    \centering
    \includegraphics[width=\linewidth]{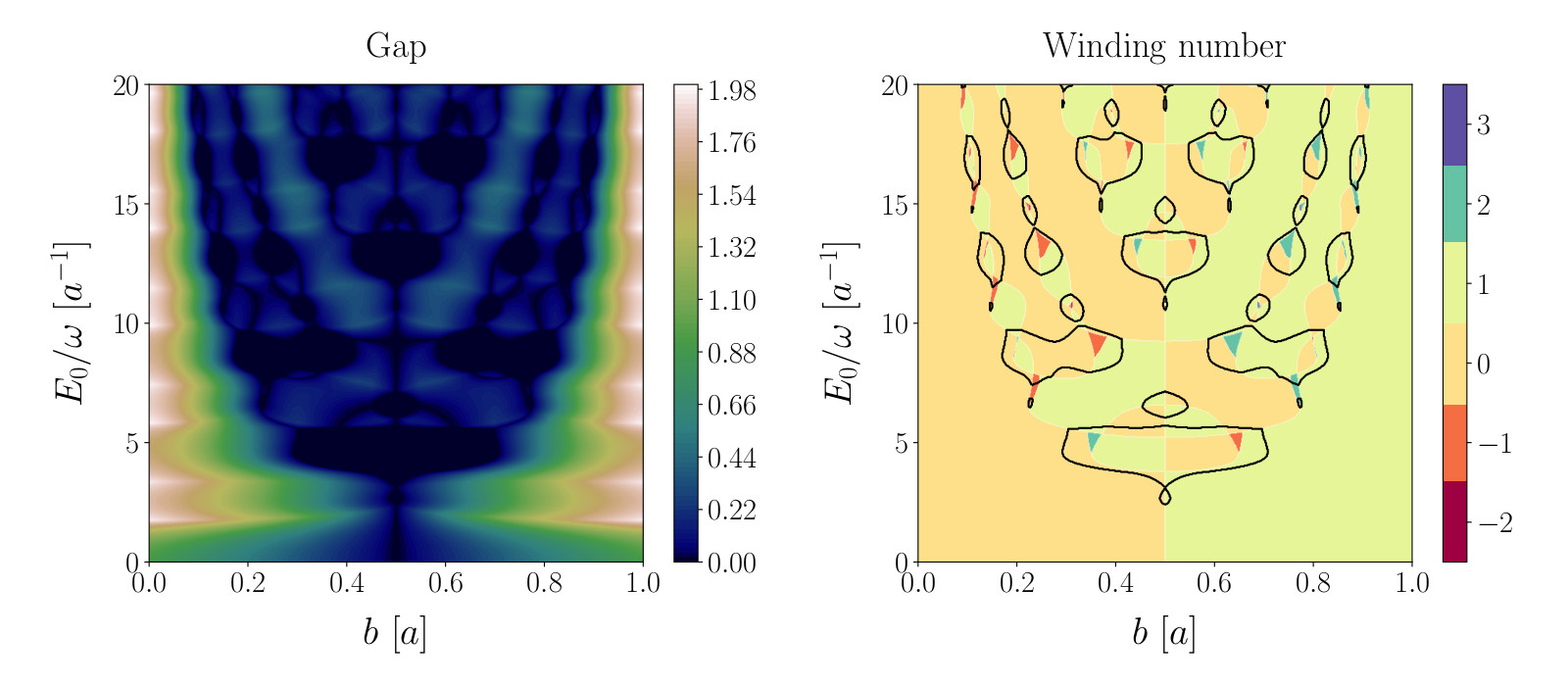}
    \caption{Phase diagram of the extended SSH model with a double sinusoidal driving $A(t)=-\frac{E_0}{w}[\sin(wt)+\sin(3w)]$. Hopping amplitudes decay exponentially with distance. Chosing $\lambda=a$, only hoppings with range up to $N=10$ have a significant contribution. The energy gap is expressed in
units of $J_0$.}
    \label{fig:PhaseDiagram_driven_super}
\end{figure}

We can also consider more complicated drivings, such as a combination of two
sinusoids with commensurate frequencies 
$E(t) = E_0\left[\cos(\omega t) + \cos(3\omega t) \right]$. 
As it can be seen in figure \ref{fig:PhaseDiagram_driven_super}, with this
driving we are able to produce gaped phases with winding numbers larger than 1, 
although the gap is smaller than in phases with smaller winding number.

An appealing option is to use a square drive. As we show below, with this kind 
of driving it is possible to zero out all even hoppings simultaneously. Let us 
consider 
\begin{equation}
    E(t) = \left\{ \begin{array}{l l l} E_0 & \mathrm{if} & 0<t<T/2 \\ 
    -E_0 & \mathrm{if} &  T/2<t<T \end{array}\right. \,,
\end{equation}
which leads to a renormalization function $f$ whose zeros are evenly spaced
on the possitive real axis, see figure \ref{fig:renormalizations} and table
\ref{tab:renormalizations}. Since the distances for all even hoppings are 
multiples of the lattice parameter $a$, it is now possible to cancel all of them 
by tuning $E_0 /\omega = 2 a^{-1}$. In this way, we can enforce chiral symmetry on
a system with arbitrarily long-range hopping terms. Despite this, with this kind 
of driving it is not possible to obtain winding numbers larger than 1 if the 
bare hopping amplitudes decay exponentially with distance. 

\section{Disorder}\label{sec:disorder}
\begin{figure}
	\centering
	\includegraphics[scale=0.7]{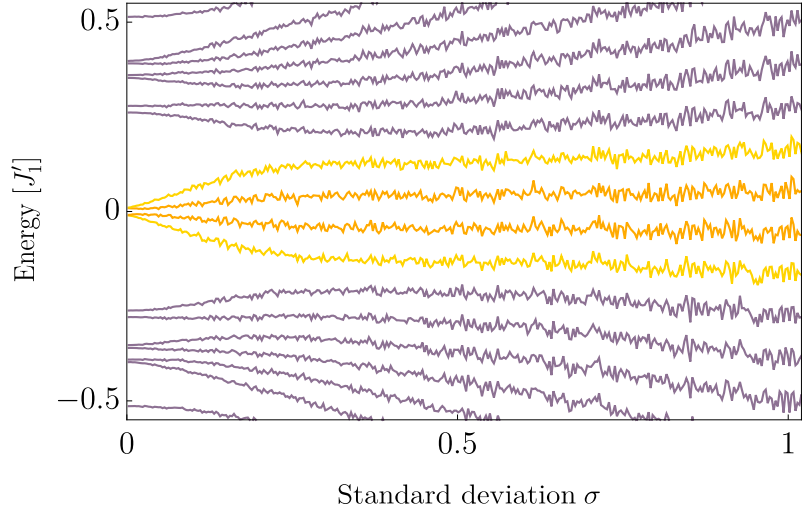}
	\caption{\label{diagonaldisorder4ES} Effect of diagonal disorder on edge states in the extended SSH model with first- and third-neighbour hoppings, as a function of the diagonal disorder strength $\sigma$. Each pair of edge states has been depicted in a different color from the states in the bands (light purple). The dimer chain has $M=20$ unit cells and hopping amplitudes $J_1=1.2J'_1$, $J_2=0$, $J'_3=0.3J'_1$, $J_3=0.9J'_1$, which are chosen such that the system has $\mathcal{W}=2$ and preserves sublattice symmetry initially ($\sigma=0$). As can be seen, the absence of sublattice symmetry separates the edge states, destroying the topological phase and leading to the usual exponential localization for arbitrary disorder strength.}
\end{figure}

The effect of disorder in electronic systems has been an important subject since Anderson's discussions on localisation
\cite{AndersonLocalization}. Originally, he studied the propagation of a particle in a random potential, and showed
that above certain critical values of disorder, localisation of the wavepackets happened. Strikingly, localisation was
extremely dependent on the spatial dimension of the system, and in 1D, they were expected to localise for infinitesimal
disorder \cite{DisorderRG}. Further studies have shown that there are exceptions to localisation in low dimensions,
being the random dimer model one of the most well known cases, where inversion symmetry leads $\sqrt{N}$ states which
are delocalised and contribute to the conductivity (i.e., they do not have zero measure in the thermodynamical
limit \cite{RandomDimerModel}). More recent studies, including its effect on the
topological phases
\cite{DisorderedTI1,DisorderedTI2,TopologicalAndersonInsulator}, and on the role
played by off-diagonal disorder
\cite{Off-diagonal-Disorder1D,Off-diagonal-Disorder,DisorderBipartiteLattices} have also been done. \\
In this section we numerically study the effect of diagonal and off-diagonal uncorrelated disorder (in the onsite energies and hopping amplitudes, respectively) in the
spectrum of both the standard and the extended SSH model. It must be stressed that this is different to the previously
mentioned random dimer model, where disorder forms a random bipartite lattice with homogeneous hoppings.\\
First, we study diagonal disorder by considering the following Hamiltonian
\begin{equation}
H'=H_\mathrm{N}+H_{\mathrm{diag}}=H_\mathrm{N}+\sum_{j=1}^{2M}\epsilon_j c^{\dagger}_jc_j\,,
\end{equation}
where the second term shifts the onsite energies differently for each site by an amount $\epsilon_j$. We use random numbers following a Gaussian distribution centered at zero, so it is necessary to average the results over several repetitions. The figures included in this section have been obtained by taking the average over 100 repetitions. Diagonal disorder breaks sublattice symmetry and eliminates the zero-energy modes, therefore destroying the topological protection of the edge modes, both in the standard and the extended SSH model (see figure \ref{diagonaldisorder4ES}).

On the other hand, off-diagonal disorder refers to random hopping amplitudes,
\begin{equation}
 H'' =H_{\mathrm{N}}+H_{\mathrm{off-diag}}=H_{\mathrm{N}}+ \sum_{|i-j|\leq N}\epsilon_{ij}c^\dagger_i c_j+ \mathrm{H.c.} \,.
\end{equation}
As it was shown in \cite{DisorderBipartiteLattices}, systems with bipartite lattices display anomalous behaviour when off-diagonal disorder is considered. One reason for this is the presence of zero energy modes at the band centre. These states appear
when sites in one sublattice couple only to sites of the other one, which is related to the differences observed in the
previous section between the effect of adding even and odd neighbour hopping. Importantly, they showed that this
type of disorder produces, at large distance and for states at $E = 0$, slow decaying localisation of the form $\propto e^{-\lambda\sqrt{r}}$ (which produces a slower random walk behaviour than the usual exponential $e^{-\lambda r}$). Debate about whether these
states are truly localised or not can be found in the literature \cite{Off-diagonal-Disorder}.

\begin{figure}[h]
	\centering
	\includegraphics[scale=0.5]{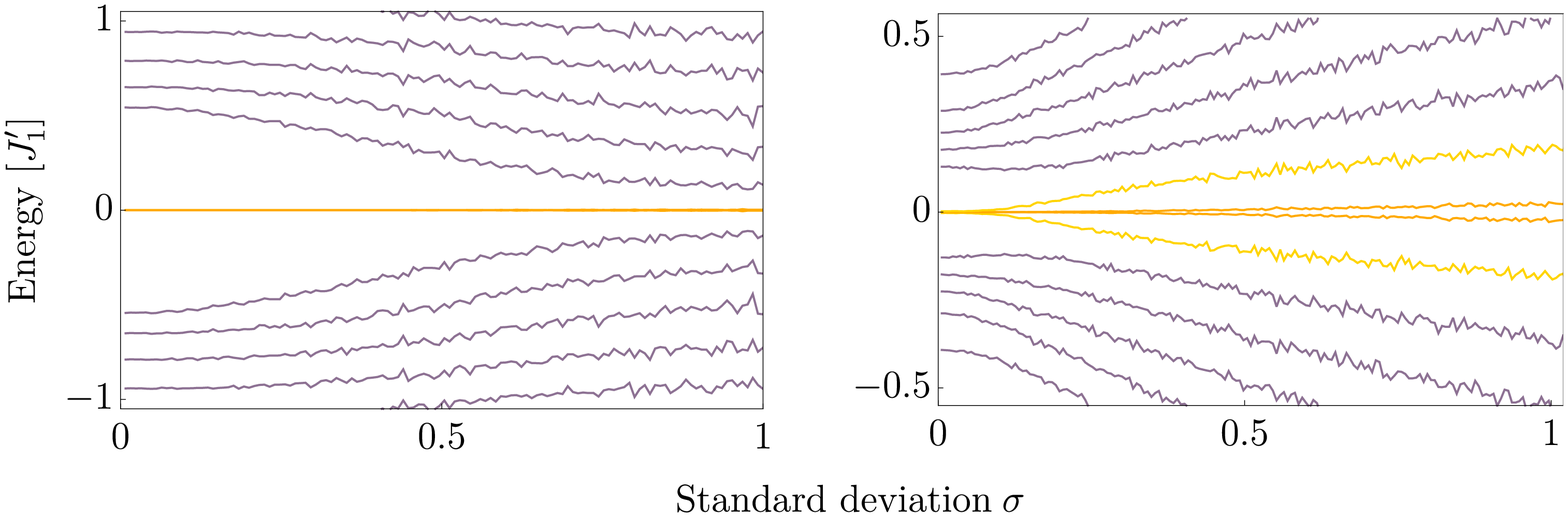}
	\caption{\label{fig:offdiagonal}Effect of off-diagonal disorder on edge states in the standard and extended SSH model with first- and third-neighbour hoppings. Each pair of edge states has been depicted in a different color from the states in the bands (light purple).\\
		(a) SSH model: finite chain with $M=20$ and $J_1=1.5J'_1$ , as a function of the off-diagonal disorder strength $\sigma$.\\ %Initially, the system has $\mathcal{W}=1$. As expected, the pair of zero-energy modes is robust under this type of perturbation, until disorder is of the order of the gap $\sigma \propto \Delta$. Then, the intra- and inter-dimer hopping cannot be differentiated, the bands mix and the edge modes separate.\\
		(b) Extended SSH model: finite chain with $M=20$ and hoppings: $J_1=1.5J'_1$, $J_2=0$, $J'_3=0.3J'_1$, $J_3=0.9J'_1$, as a function of the off-diagonal disorder strength $\sigma$. Hoppings are chosen such that the system has ($\mathcal{W}=2$) initially.} %It is interesting to see that in this case, each pair of edge states behaves differently under the increase of disorder. }
\end{figure}

Figure \ref{fig:offdiagonal} shows how off-diagonal disorder affects edge states in both the standard and extended SSH model for a configuration with $\mathcal{W}=2$ and first- to third- neighbour hoppings. As expected, the pair of zero-energy modes in a SSH chain is robust under this type of perturbation, until disorder is of the order of the gap $\sigma \propto \Delta$. Then, the intra- and inter-dimer hopping cannot be differentiated, the bands mix and eventually the edge modes separate. However, it is interesting to see how each pair of edge states behaves differently when disorder is increased in the extended-SSH configuration under consideration.

\section{Conclusions\label{sec:conclusions}}
%Long range interactions occur in several physical systems. In particular, polyacetilene, a dimer chain,  has been modelled by the cellebrated SSH model which includes just first neighbors. 
In this work, we have studied a generalized model for a dimer chain including
long-range hoppings, which naturally occur in physical systems. Although seemingly equal, the effect of hopping processes
connecting the same sublattice (even hoppings), and processes connecting
different sublattices (odd hoppings) is very different. The former breaks
particle-hole symmetry, and changes the topological class from BDI to AI.
Nevertheless, the presence of space inversion symmetry forces the topological
invariant to have quantized values, and the appearance of edge states protected
only by this symmetry.  As a consequence, the number of edge states now changes
independently of the topological invariant, as they can enter the bulk bands if
the hopping amplitudes connecting different sublattices is large enough. On the
contrary, hopping between different sublattices preserves the fundamental
symmetries, and allows for phases with larger values of the topological
invariant and larger numbers of edge-state pairs.

We propose the use of an ac driving to tune the topological properties of the system. Three different drivings are analyzed. Interestingly, we show that with a square-wave driving it is possible to cancel all even hoppings simultaneously, restoring the symmetries of the standard SSH model.

Finally, we have investigated the effect of disorder. In the case of a chain with only odd hoppings, the edge states are robust against off-diagonal disorder, while they loose their protection as long as we introduce even hoppings. We also show that in phases with more than a single pair of edge states, their energies departure from zero at different rates as the strength of diagonal disorder is increased.

\section*{Acknowledgements}
This work was supported by the Spanish Ministry of
Economy and Competitiveness through Grant No.MAT2014-58241-P. M. Bello acknowledges the FPI program (BES-2015-071573), and A. G\'omez-Le\'on acknowledges the Juan de la Cierva program.

\appendix

\section{Topological invariant in 1D systems \label{appendixZ}}

For 1D systems, we can define a topological invariant through the Zak phase \cite{zakphase1989}, which is the integration of the Berry connection over the first Brillouin zone (FBZ).
\begin{equation}
\mathcal{Z}_n=i\int_{\mathrm{FBZ}} dk \mathcal{h} u_n(k)|\partial_k u_n(k)\mathcal{i}\,,
\label{zakphase}
\end{equation}
where $n$ is the band subscript ($n=\pm$) and $|u_n(k)\mathcal{i}$ are the Bloch functions.
The Zak phase is a particular case of the Berry phase \cite{berryphase1984}, which is the geometric phase acquired by an eigenstate of the system when it is made to evolve cyclically in the parameter space of the problem under consideration. When this concept is applied to the dynamics of electrons in periodic solids, the Berry phase is refered to as the Zak phase and the parameter space, the Brillouin zone, is naturally furnished by the system itself. When $k$ is sweeped across the FBZ $k=0\rightarrow 2\pi$, eigenstates evolves through a closed path, picking up a phase given by \eref{zakphase}.\\
The Zak phase is closely related to the bulk electric polarization, as has been shown in the so-called modern theory of polarization . The bulk electric polarization is given by $P_{\mathrm{bulk}}=\mathcal{Z}/2\pi$. In a neutral chain, $P_{\mathrm{total}}=P_{\mathrm{edge}}+P_{\mathrm{bulk}}=0$. If the bulk polarization is non-zero, there must be accumulation of charge at the edges, thus explaining the relation between a non-zero value of the Zak phase and the presence of edge states. \\
On the other hand, Bloch functions are eigenstates of the bulk momentum-space Hamiltonian, $\mathcal{H}|u_n(k)\mathcal{i}=E_n(k)|u_n(k)\mathcal{i}$. The Zak phase can be understood as the rotation angle $|u_n(k)\mathcal{i}$ undergoes when it is parallel transported along the FBZ. The curvature of the FBZ, reflected in the Berry connection, is responsible for the phase that the Bloch function picks up. 

Equivalently, it can be expressed in terms of the winding of the closed curve defined by the Bloch vector as $k=0\rightarrow 2\pi$ around the origin, 
\begin{equation}
\W=\frac{\mathcal{Z}}{\pi}=\frac{1}{2\pi}\int_{\mathrm{FBZ}}\frac{d_x \partial_k d_y-d_y \partial_k d_x}{d_x^2+d_y^2}dk \,.
\end{equation}
In the SSH model, the curve $\gamma=(d_x(k),d_y(k))=(J'_1+J_1\cos(k),J_1\sin(k))$ describes a circunference centered at $(J'_1,0)$ and radius $J_1$. Thus, the topology of a system is determined by whether or not the previous curve encloses the origin. In this geometric picture, we can identify topologically equivalent configurations as those whose $\gamma$ can be continuously transformed into one another without passing through the origin. In the extended SSH model, the topological phase diagram is enriched and $\gamma$ displays more complex geometries, giving raise to larger values of the winding. \\
%The Zak phase is a gauge invariant quantity and as such can be observed. It was recently measured in ultra-cold atoms \cite{measurementZak}. Apart from the SSH model of polyacetylene \cite{ssh1980}, the Zak phase can also be found in linearly conjugated diatomic polymers \cite{conjugatedpolymers}, photonic systems \cite{photonic1,photonic2}, acoustic systems \cite{acousticsystems}, and recently, water wave states \cite{topologicalwater}. 

\section{Floquet theory}\label{app:Floquet} 
The starting Hamiltonian is $H(t) = H_N + H_{AC}(t)$.  
For a time-periodic Hamiltonian, $H(t+T)=H(t)$ with $T=2\pi/\omega$, Floquet's theorem permits us to write the time-evolution operator $U(t_2,t_1)$ as
\begin{equation}
	U(t_2,t_1)=e^{-iK(t_2)}e^{-i\Heff(t_2-t_1)}e^{iK(t_1)} \,,
\end{equation}
where $\Heff$ is a time independent (effective) Hamiltonian and $K(t)$ is a 
$T$-periodic Hermitian operator. $\Heff$ governs the long-term dynamics whereas 
$e^{-iK(t)}$, also known as the \textit{micromotion operator}, accounts for the 
fast dynamics occurring within a period. Following several perturbative methods 
\cite{Eckardt2015,Bukov2015}, it is possible to find expressions for these 
operators as power series in $1/\omega$
\begin{equation}
	\Heff=\sum\limits_{n=0}^\infty \frac{H^{[n]}}{\omega^n} \,, \quad 
	K(t)=\sum\limits_{n=0}^\infty \frac{K^{[n]}(t)}{\omega^n} \,.
\end{equation}
The different terms in these expansions have a progressively more complicated 
dependence on the Fourier components of the original Hamiltonian, 
${H^{(q)}=T^{-1}\int_0^T H(t)e^{i\omega q t}dt}$. The 
first three of them are:
\begin{eqnarray}
	H^{[0]} = H^{(0)} \,, \quad
	H^{[1]} = \sum\limits_{q\neq 0}\frac{H^{(-q)}H^{(q)}}{q} \,, \\
	H^{[2]} = \sum\limits_{q, p\neq 0}\left(
	\frac{H^{(-q)}H^{(q-p)}H^{(p )}}{q p} -
	\frac{H^{(-q)}H^{(q)}H^{(0)}}{q^2} \right) \,.
\end{eqnarray}

Before deriving the effective Hamiltonian, in order to obtain a result that is
non-perturbative in the ac field amplitude, it is convenient to transform the 
original Hamiltonian into the rotating frame with respect to the ac field
\begin{eqnarray}
    \Hrot(t)=\U^\dagger(t)H(t)\U(t)
    -i\U^\dagger(t)\partial_t\U(t) \,, \\
	\U(t)=e^{-i\int H_{AC}(t)dt} \,.
\end{eqnarray}
This leads to
\begin{equation}
    \Hrot(t) = \sum_{i,j} J_{ij}e^{iA(t)d_{ij}} c^\dagger_i c_j \,.
\end{equation}

\section*{References}

\bibliographystyle{iopart-num}
\bibliography{floquet,longrangeSSH}

\providecommand{\newblock}{}
\begin{thebibliography}{10}
\expandafter\ifx\csname url\endcsname\relax
  \def\url#1{{\tt #1}}\fi
\expandafter\ifx\csname urlprefix\endcsname\relax\def\urlprefix{URL }\fi
\providecommand{\eprint}[2][]{\url{#2}}

\bibitem{qhe1982}
Thouless D~J, Kohmoto M, Nightingale M~P and den Nijs M 1982 {\em Phys. Rev.
  Lett.\/} {\bf 49}(6) 405--408
  \urlprefix\url{https://link.aps.org/doi/10.1103/PhysRevLett.49.405}

\bibitem{fqhe1983}
Laughlin R~B 1983 {\em Phys. Rev. Lett.\/} {\bf 50}(18) 1395--1398
  \urlprefix\url{https://link.aps.org/doi/10.1103/PhysRevLett.50.1395}

\bibitem{sphi2007}
K{\"o}nig M, Wiedmann S, Br{\"u}ne C, Roth A, Buhmann H, Molenkamp L~W, Qi X~L
  and Zhang S~C 2007  {\bf 318} 766--770

\bibitem{spintopology2014}
Cinchetti M 2014 {\em Nature Nanotechnology\/} {\bf 9} 965
  \urlprefix\url{http://dx.doi.org/10.1038/nnano.2014.284}

\bibitem{magtopology2013}
Nagaosa T 2013 {\em Nature Nanotechnology\/} {\bf 8} 899--911
  \urlprefix\url{http://dx.doi.org/10.1038/nnano.2013.243}

\bibitem{qckitaev2003}
Kitaev A 2003 {\em Annals of Physics\/} {\bf 303} 2 -- 30 ISSN 0003-4916
  \urlprefix\url{http://www.sciencedirect.com/science/article/pii/S0003491602000180}

\bibitem{majoranabox2017}
Plugge S, Rasmussen A, Egger R and Flensberg K 2017 {\em New Journal of
  Physics\/} {\bf 19} 012001
  \urlprefix\url{http://stacks.iop.org/1367-2630/19/i=1/a=012001}

\bibitem{firstthirdneighbours}
Li L, Yang C and Chen S 2015 {\em EPL (Europhysics Letters)\/} {\bf 112} 10004
  \urlprefix\url{http://stacks.iop.org/0295-5075/112/i=1/a=10004}

\bibitem{generalizedSSH}
Li L, Xu Z and Chen S 2014 {\em Phys. Rev. B\/} {\bf 89}(8) 085111
  \urlprefix\url{https://link.aps.org/doi/10.1103/PhysRevB.89.085111}

\bibitem{ssh1980}
Su W~P, Schrieffer J~R and Heeger A~J 1980 {\em Phys. Rev. B\/} {\bf 22}(4)
  2099--2111 \urlprefix\url{https://link.aps.org/doi/10.1103/PhysRevB.22.2099}

\bibitem{continuumSSH1980}
Takayama H, Lin-Liu Y~R and Maki K 1980 {\em Phys. Rev. B\/} {\bf 21}(6)
  2388--2393 \urlprefix\url{https://link.aps.org/doi/10.1103/PhysRevB.21.2388}

\bibitem{hubbardvspeierls}
Kivelson S and Heim D~E 1982 {\em Phys. Rev. B\/} {\bf 26}(8) 4278--4292
  \urlprefix\url{https://link.aps.org/doi/10.1103/PhysRevB.26.4278}

\bibitem{measurementZak}
Marcos~Atala Monika~Aidelsburger J~T~B~D~A~T~K~E~D and Bloch I 2013 {\em Nature
  Physics\/} {\bf 9} 795–800

\bibitem{conjugatedpolymers}
Rice M~J and Mele E~J 1982 {\em Phys. Rev. Lett.\/} {\bf 49}(19) 1455--1459
  \urlprefix\url{https://link.aps.org/doi/10.1103/PhysRevLett.49.1455}

\bibitem{photonic1}
Longhi S 2013 {\em Opt. Lett.\/} {\bf 38} 3716--3719
  \urlprefix\url{http://ol.osa.org/abstract.cfm?URI=ol-38-19-3716}

\bibitem{photonic2}
Wei~Tan Yong~Sun H~C~S~Q~S 2014 {\em Scientific Reports\/} {\bf 4} 3842

\bibitem{acousticsystems}
Meng~Xiao Guancong~Ma Z~Y~P~S~Z~Q~Z~~C~T~C 2014 {\em Nature Physics\/} {\bf 11}
  240--244

\bibitem{topologicalwater}
Zhaoju~Yang F~G~~B~Z 2016 {\em Scientific Reports\/} {\bf 6} 29202

\bibitem{tenfoldway}
Ryu S, Schnyder A~P, Furusaki A and Ludwig A~W~W 2010 {\em New Journal of
  Physics\/} {\bf 12} 065010
  \urlprefix\url{http://stacks.iop.org/1367-2630/12/i=6/a=065010}

\bibitem{asboth2016book}
Asb{\'o}th J, Oroszl{\'a}ny L and P{\'a}lyi A 2016 {\em A Short Course on
  Topological Insulators: Band Structure and Edge States in One and Two
  Dimensions\/} Lecture Notes in Physics (Springer International Publishing)
  ISBN 9783319256078
  \urlprefix\url{https://books.google.es/books?id=RWKhCwAAQBAJ}

\bibitem{delplace2011}
Delplace P, Ullmo D and Montambaux G 2011 {\em Phys. Rev. B\/} {\bf 84}(19)
  195452 \urlprefix\url{https://link.aps.org/doi/10.1103/PhysRevB.84.195452}

\bibitem{drivenchain2013}
G\'omez-Le\'on A and Platero G 2013 {\em Phys. Rev. Lett.\/} {\bf 110}(20)
  200403
  \urlprefix\url{https://link.aps.org/doi/10.1103/PhysRevLett.110.200403}

\bibitem{Hanggi1998}
Grifoni M and H\"anggi P 1998 {\em Physics Reports\/} {\bf 304} 229 -- 354 ISSN
  0370-1573
  \urlprefix\url{http://www.sciencedirect.com/science/article/pii/S0370157398000222}

\bibitem{AndersonLocalization}
Anderson P~W 1958 {\em Phys. Rev.\/} {\bf 109}(5) 1492--1505
  \urlprefix\url{https://link.aps.org/doi/10.1103/PhysRev.109.1492}

\bibitem{DisorderRG}
Abrahams E, Anderson P~W, Licciardello D~C and Ramakrishnan T~V 1979 {\em Phys.
  Rev. Lett.\/} {\bf 42}(10) 673--676
  \urlprefix\url{https://link.aps.org/doi/10.1103/PhysRevLett.42.673}

\bibitem{RandomDimerModel}
Dunlap D~H, Wu H~L and Phillips P~W 1990 {\em Phys. Rev. Lett.\/} {\bf 65}(1)
  88--91 \urlprefix\url{https://link.aps.org/doi/10.1103/PhysRevLett.65.88}

\bibitem{DisorderedTI1}
Bardarson J~H, Brouwer P~W and Moore J~E 2010 {\em Phys. Rev. Lett.\/} {\bf
  105}(15) 156803
  \urlprefix\url{https://link.aps.org/doi/10.1103/PhysRevLett.105.156803}

\bibitem{DisorderedTI2}
Zhang X, Guo H and Feng S 2012 {\em Journal of Physics: Conference Series\/}
  {\bf 400} 042078
  \urlprefix\url{http://stacks.iop.org/1742-6596/400/i=4/a=042078}

\bibitem{TopologicalAndersonInsulator}
Li J, Chu R~L, Jain J~K and Shen S~Q 2009 {\em Phys. Rev. Lett.\/} {\bf
  102}(13) 136806
  \urlprefix\url{https://link.aps.org/doi/10.1103/PhysRevLett.102.136806}

\bibitem{Off-diagonal-Disorder1D}
Pendry J~B 1982 {\em Journal of Physics C: Solid State Physics\/} {\bf 15} 5773
  \urlprefix\url{http://stacks.iop.org/0022-3719/15/i=28/a=009}

\bibitem{Off-diagonal-Disorder}
Biswas P, Cain P, Römer R and Schreiber M 2000 {\em physica status solidi
  (b)\/} {\bf 218} 205--209 ISSN 1521-3951
  \urlprefix\url{http://dx.doi.org/10.1002/(SICI)1521-3951(200003)218:1<205::AID-PSSB205>3.0.CO;2-B}

\bibitem{DisorderBipartiteLattices}
Inui M, Trugman S~A and Abrahams E 1994 {\em Phys. Rev. B\/} {\bf 49}(5)
  3190--3196 \urlprefix\url{https://link.aps.org/doi/10.1103/PhysRevB.49.3190}

\bibitem{zakphase1989}
Zak J 1989 {\em Phys. Rev. Lett.\/} {\bf 62}(23) 2747--2750
  \urlprefix\url{https://link.aps.org/doi/10.1103/PhysRevLett.62.2747}

\bibitem{berryphase1984}
V~Berry M 1984  {\bf 392} 45--57

\bibitem{Eckardt2015}
Eckardt A and Anisimovas E 2015 {\em New J. of Phys.\/} {\bf 17} 093039
  \urlprefix\url{http://stacks.iop.org/1367-2630/17/i=9/a=093039}

\bibitem{Bukov2015}
Bukov M, D'Alessio L and Polkovnikov A 2015 {\em Adv. in Phys., vol. 64, No.
  2\/}  139--226

\end{thebibliography}

\end{document}